\documentclass[twocolumn]{aastex631}

\usepackage{graphicx}	
\usepackage{threeparttable} 
\usepackage[utf8]{inputenc}
\usepackage[T1]{fontenc}
\usepackage{amssymb}
\usepackage{amsmath}
\usepackage{hyperref}
\usepackage{xspace} 

\newcommand{\ixpe}{IXPE\xspace}
\newcommand{\nustar}{NuSTAR\xspace}
\newcommand{\nicer}{NICER\xspace}
\newcommand{\swift}{{Swift}\xspace}
\newcommand{\kerrbb}{\texttt{kerrbb}\xspace}
\newcommand{\kynbbrr}{\texttt{kynbbrr}\xspace}
\newcommand{\kynebbrr}{\texttt{kynebbrr}\xspace}

\begin{document}

%\title{X-ray polarization confirms low black-hole spin and high inclination in LMC~X-3}
\title{First X-ray polarization measurement confirms the low black-hole spin in \mbox{LMC~X-3}}

%\author{IXPE Colls...}
%\affiliation{All affiliations listed at the end of the paper.}

%this file can be used to make the list of authors
% currently it includes only tier-2 authors in alphabetical order
\author[0000-0003-2931-0742]{Ji\v{r}\'{i} Svoboda\thanks{E-mail: jiri.svoboda@asu.cas.cz}}
\affiliation{Astronomical Institute of the Czech Academy of Sciences, Bo\v{c}n\'{i} II 1401/1, 14100 Praha 4, Czech Republic}

\author[0000-0003-0079-1239]{Michal Dov\v{c}iak}
\affiliation{Astronomical Institute of the Czech Academy of Sciences, Bo\v{c}n\'{i} II 1401/1, 14100 Praha 4, Czech Republic}

\author[0000-0002-5872-6061]{James F. Steiner}
\affiliation{Center for Astrophysics, Harvard \& Smithsonian, 60 Garden St, Cambridge, MA 02138, USA}

\author[0000-0003-3331-3794]{Fabio Muleri}
\affiliation{INAF Istituto di Astrofisica e Planetologia Spaziali, Via del Fosso del Cavaliere 100, 00133 Roma, Italy}	

\author[0000-0002-5311-9078]{Adam Ingram}
\affiliation{School of Mathematics, Statistics, and Physics, Newcastle University, Newcastle upon Tyne NE1 7RU, UK}

\author[0000-0003-1133-1684]{Anastasiya Yilmaz}
\affiliation{Astronomical Institute of the Czech Academy of Sciences, Bo\v{c}n\'{i} II 1401/1, 14100 Praha 4, Czech Republic}
\affiliation{Astronomical Institute, Faculty of Mathematics and Physics, Charles University, V Holešovičkách 2, Prague 8, 180~00, Czech Republic}
\affiliation{Institute of Theoretical Physics, Faculty of Mathematics and Physics, Charles University, V Holešovickách 2, CZ-180 00 Praha 8, Czech Republic}

\author[0000-0001-5256-0278]{Nicole Rodriguez Cavero}
\affiliation{Physics Department, McDonnell Center for the Space Sciences, and Center for Quantum Leaps, Washington University in St. Louis, St. Louis, MO 63130, USA}

\author[0009-0001-4644-194X]{Lorenzo Marra}
\affiliation{Dipartimento di Matematica e Fisica, Universit\`a degli Studi Roma Tre, Via della Vasca Navale 84, 00146 Roma, Italy}

\author[0000-0002-0983-0049]{Juri Poutanen}
\affiliation{Department of Physics and Astronomy,  20014 University of Turku, Finland}

\author[0000-0002-5767-7253]{Alexandra Veledina}
\affiliation{Department of Physics and Astronomy, 20014 University of Turku, Finland}
\affiliation{Nordita, KTH Royal Institute of Technology and Stockholm
University, Hannes Alfv\'ens v\"ag 12, SE-10691 Stockholm, Sweden}

% ...

\begin{abstract}
X-ray polarization is a powerful tool to investigate the geometry of accreting material around black holes, allowing independent measurements of the black hole spin and orientation of the innermost parts of the accretion disk. We perform the X-ray spectro-polarimetric analysis of an X-ray binary system in the Large Magellanic Cloud, \mbox{LMC~X-3}, that hosts a stellar-mass black hole, known to be persistently accreting since its discovery. We report the first detection of the X-ray polarization in \mbox{LMC~X-3} with the Imaging X-ray Polarimetry Explorer,
%used to measure the X-ray polarization, accompanied by multiple spectroscopic observations using the NuSTAR, NICER, and the Neil Gehrels Swift observatories.
% The multi-instrument spectral analysis prefers a slim disk model to fit the X-ray spectra. 
and find the average polarization degree of $3.2\% \pm 0.6 \%$ and a constant polarization angle $-42\degr \pm 6\degr$ over the 2--8\,keV range. Using accompanying spectroscopic observations by NICER, NuSTAR, and the Neil Gehrels Swift observatories, we confirm previous measurements of the black hole spin via the X-ray continuum method, $a \approx 0.2$. From polarization analysis only, we found consistent results with low black-hole spin, with an upper limit of $a < 0.7$ at a 90\% confidence level. 
%
%), and the accretion disk has an inclination $i\approx70\degr$ consistently with the binary orientation.
%, which are consistent with the previous estimates. 
%The independent $Q/I$ and $U/I$ Stokes spectra constrain the black hole spin $a/M < 0.66$ with no returning radiation assumed, and $a/M < 0.3$ with ideally reflecting returning radiation, respectively.
%Our spectropolarimetric modeling suggests the black hole in this binary has low spin $a<0.66$ and the binary inclination is $i\approx70\degr$, which are consistent with the previous estimates. 
%are consistent with a low value of the black-hole spin $\approx 0.2$ and high inclination angle $\approx 70$\,deg,  constrained by the spectroscopic X-ray continuum method. 
A slight increase of the polarization degree with energy, similar to other black-hole X-ray binaries in the soft state, is suggested from the data but with a low statistical significance. 
\end{abstract}


\section{Introduction} \label{Introduction}

Accreting stellar-mass black hole X-ray binaries (BHXRBs) are among the brightest X-ray sources in our Galaxy.
%and are thus one of the most suitable targets for the recently launched X-ray polarization mission {Imaging X-ray Polarimetry Explorer} (\ixpe; \citet{Weisskopf2022}).
%The X-ray polarization signal carries signatures of the geometry and nature of the origin of the X-ray emission. 
%Their X-ray emission carries information about the accretion processes and the central object. In particular, the black hole mass and spin are important parameters that determine the spacetime around the compact object. 
Only a few are persistent sources.
%, which have been in an outburst since the beginning of the X-ray observations.
Most of them are transients characterized by short, weeks to months, periods of activity and longer, years to decades, quiescent episodes \citep{FKR2002, Zdziarski2004}.
BHXRBs have been found to swing between different spectral states, which are distinguished by broadband spectra and timing characteristics \citep{Zdziarski2004,Done2007,Belloni2010}.
In the hard state, the spectrum has a power-law-like shape that is believed to be produced by multiple Compton (up-)scattering events of low-energy photons in a hot medium, referred to as a hot accretion flow or a corona.
%The geometry and location of this medium is debated, along with the possibility of extracting the parameters of the central BHs \citep{Done2007,Bambi2021}.
%
The soft-state spectrum resembles blackbody radiation, and is commonly attributed to the multicolor emission of the disk \citep{Shakura1973, Novikov1973}, which is extending down to the innermost stable circular orbit (ISCO).
The shape of the spectrum is tightly related to the radius of the ISCO and, by extension, the black-hole spin.
This property enables the determination of BH spins using the so-called continuum fitting method \citep{Shafee2006,McClintock2014}.

X-ray polarization offers an alternative means to probe the topology of accreting matter and these measurements have become accessible following the launch of the Imaging X-ray Polarimetry Explorer \citep[IXPE,][]{Weisskopf2022}.
The first results on the hard-state BHXRB \mbox{Cyg~X-1} suggest the hot medium is extended along the disc plane \citep{Krawczynski2022}, and may have a substantial outflow velocity \citep{Poutanen2023}.
Another persistent source \mbox{Cyg~X-3} instead was found to possess an optically thick, elevated envelope resulting from super-Eddington accretion \citep{Veledina2023}.

It has been anticipated that the energy dependence of X-ray polarization degree (PD) and polarization angle (PA) in the soft-state sources could be utilized for measuring the black-hole spin \citep[e.g.,][]{Dovciak2008, Schnittman2009, Mikusincova2023}.
Polarization in the accretion disk \citep{Rees1975} can be produced by electron scattering in the upper layers of the disk (a disk atmosphere).
For the case of a razor-thin semi-infinite atmosphere, the PD is a known function of the disk inclination \citep{Chandrasekhar1960, Sobolev1963}.
Light bending, relativistic aberration and frame-dragging effects modify the viewing angles of different parts of the accretion disk, resulting in a characteristic rotation of PA and an overall depolarization, both of which depend on the BH spin \citep{ConnorsStark1977, Connors1980}.

Yet, the first polarization measurements of a BHXRB in the soft-state, \mbox{4U~1630$-$47}, revealed severe problems with this standard scenario, as neither a characteristic de-polarization nor PA rotation have been observed \citep{Ratheesh2023}.
On the contrary, the observed high PD exceeding 8\% and its increase with energy pose significant challenges to all existing models.
These models thus invariably require parameter adjustments to account for the observed values.
Follow-up observation of \mbox{4U~1630$-$47} in the steep power-law state (when both the disk and Comptonization continuum significantly contribute to the X-ray band) emphasize the difficulties highlighted by the soft-state data \citep{Rodriguez-Cavero2023}.

Other \ixpe observations of BHXRBs in the soft state include \mbox{LMC~X-1} \citep{Podgorny2023}, \mbox{4U~1957+115} (Marra et al. in prep.), and \mbox{Cyg~X-1} (\citet{Dovciak2023}, Steiner et al. in prep.). LMC~X-1 is a low-inclination system. As expected, it has exhibited a low PD around 1\% \citep{Podgorny2023}. In addition, a significant contribution to the polarization was possibly due to a Comptonization component. A higher inclination system, 4U~1957+115, revealed a higher polarization fraction of PD\,$\approx 2\%$ (Marra et al., in prep.). Overall, the polarization properties of both sources were found to be consistent with theoretical expectations given their inclinations.

%In this work, we report the first X-ray polarization measurement of a persistent X-ray binary \mbox{LMC~X-3}, located in the Large Magellanic Cloud.
The first X-ray polarization measurement of \mbox{LMC~X-3} is the subject of the present study. \mbox{LMC~X-3} is an X-ray binary located in the Large Magellanic Cloud (LMC) at the most recently estimated distance $D = 49.59 \pm 0.09$ (statistical) $\pm 0.54$ (systematic) kpc \citep{Pietrzynski2019}.
The mass of the black hole, companion star, and the inclination of the system are constrained from optical photometric and spectroscopic observations: $M_{\rm BH} = 6.98\pm0.56 M_{\odot}$, $M_{\rm star} = 3.63\pm0.57 M_{\odot}$, and $i = 69.2\degr$ \citep{Orosz2014}, where $M_{\odot}$ is the mass of the Sun. 
In the X-rays, \mbox{LMC~X-3} was first detected by {UHURU} satellite \citep{Leong1971} and has subsequently been observed by all major X-ray satellites, owing to its persistent nature.
These studies revealed the source to reside primarily in the soft state \citep{Treves1988, Ebisawa1993, Nowak2001}, with only rare hard-state occurences \citep{Wilms2001, Wu2001} and occasional entry into an anomalous low state characterized by a drop in X-ray flux by a few orders of magnitude \citep{Smale2012, Torpin2017}. 

The \mbox{LMC~X-3}'s soft-state spectrum was found to be well modeled by a multicolor disk, whose inner temperature is proportional to the fourth root of the X-ray luminosity \citep{Gierlinski2004}.
Given the known distance and almost persistent stay in the high/soft state, \mbox{LMC~X-3} has been identified as one of the most promising targets for black-hole spin measurements using the X-ray continuum fitting method. \citet{Steiner2010} analyzed a large set of RXTE observations and found a constant inner disk radius until reaching a critical luminosity, found to be around $0.3\,L_{\rm Edd}$ (where $L_{\rm Edd}$ is the Eddington luminosity, $L_{\rm Edd} =  1.26 \times 10^{38} {M}/{M_{\odot}}$\,erg\,s$^{-1}$).
%\footnote{The Eddington luminosity $L_{\rm Edd}$ is the maximum luminosity for which the radiation pressure acting outwards does not overcome the gravitational force acting inwards, we use the value $L_{\rm Edd} =  1.26 \times 10^{38} {M}/{M_{\odot}}$\,erg\,s$^{-1}$ calculated for pure ionized hydrogen.} 
% the standard thin Novikov-Thorne disk model \citep{Novikov1973}.
%{\textbf{}}
For higher luminosities, the measured value for the innermost disk radius increased, indicating a change in the structure of the accretion disk or the disk atmosphere. 
To account for this behavior, slim-disk models %employed for \mbox{LMC~X-3} 
were developed \citep{Straub2011}. The slim disk is a solution with the aspect ratio $H/R \lesssim 1$ (where $H$ is the scale-height of the disk and $R$ is the radius from the center). However, the apparent increase of the innermost disk radius at high luminosity remained unsolved.

The black hole spin is closely related to the innermost disk radius, assuming the accretion disk extends down to ISCO. The first spin estimates were affected by uncertainty due to the unknown mass of the black hole. More accurate spin measurements were possible only following precise determination of the black hole's mass from optical spectroscopy \citep{Orosz2014}. The spin was measured through the X-ray continuum fitting method as $a \approx 0.2$ \citep{Steiner2014}.
The low value of the black-hole spin has been subsequently confirmed in more recent analyses \citep{Bhuvana2022, Yilmaz2023}. 
\citet{Yilmaz2023} reported a measured value for black-hole spin as $a \approx 0.1$. In their analysis, they relaxed the condition of a constant innermost radius at ISCO and showed a scatter of the inner disk radius measurements in different observations during the outbursts (see their Figure 7 and 9).

In this paper, we present the analysis of the first IXPE observations of \mbox{LMC~X-3}, showing that the polarimetric data are in line with the low spin of the black hole, previously measured only by X-ray spectroscopy. The observations are described in Section~\ref{Observations}. The results of the spectro-polarimetric modeling are presented in Section~\ref{Results} and discussed in Section~\ref{Discussion}. Section~\ref{Summary} summarizes the main results of the analysis.

\begin{table*}
\centering
\footnotesize
\caption{List of observations.}
%no vertical lines are allowed! 
\begin{tabular}{cccccc}
\hline
\hline
\rule{0cm}{0.3cm}
Observatory & Instrument & Observation ID & Start Date & End Date & Net exposure [ks] \\  
\hline
%\hline
IXPE & GPD & 02006599 & 2023-07-07 18:42:31 & 2023-07-09 13:39:07 & 105.2 \\ 
 & & 02006599 & 2023-07-12 16:55:48 & 2023-07-20 14:19:34 & 458.6 \\
NICER &  XTI & 6101010117 & 2023-07-08 10:10:50 & 2023-07-08 16:34:40 & 2.14 \\
& & 6101010118 & 2023-07-17 23:37:19 & 2023-07-17 23:40:14 & 0.16 \\
NuSTAR &  FPMA/FPMB & 309020041002 & 2023-07-09 12:16:09 & 2023-07-09 23:51:09& 27.6 \\
 & & 309020041004 &2023-07-14 07:56:09 & 2023-07-15 06:31:32 & 28.1 \\
 & & 309020041006 & 2023-07-19 19:41:09 & 2023-07-21 07:16:23 & 29.0
 \\
\swift & XRT & 00089714001 & 2023-07-09 19:29:38 & 2023-07-09 19:31:50 & 1.3\\
 & & 00089714002 & 2023-07-14 02:50:08 & 2023-07-14 16:54:56 & 1.8 \\
 & & 00089714003 & 2023-07-20 01:21:15 & 2023-07-20 04:42:56 & 1.9 \\
\hline
% only one horizontal line !
%\hline
\end{tabular}
\label{table:list_observations}
\end{table*}

\section{Observations} \label{Observations}

\mbox{LMC~X-3} was observed in July 2023 by multiple X-ray instruments. All studied observations are summarized in Table~\ref{table:list_observations}. Data reduction and processing are described in more detail in the following sections.

\subsection{IXPE}

\mbox{LMC~X-3} was observed by {\ixpe} on 2023 July 7--8 and 12--21 (ObsID: 02006599) with a total exposure time 562\,ks. IXPE detectors \citep{Soffitta2021} can measure the Stokes parameters $I$, $Q$, and $U$ and they have imaging capabilities, so that the source and background regions can be spatially separated. Level 2 event files were downloaded from the HEASARC and then filtered for source and background regions using \texttt{xpselect} tool from the \textsc{ixpeobssim} software package version 30.5 \citep{Baldini2022}. 
%We used \textsc{ixpeobssim} version 30.5 for the data reduction.  
The source extraction regions were selected for each detector unit as circles with a radius of 60\arcsec.
The background regions were defined as annuli with an inner radius of 180\arcsec\ and the outer radius of 280\arcsec.

Polarization cubes were generated using the unweighted \texttt{pcube} algorithm \citep{Baldini2022}. We produced PD and PA in 5 energy bins and obtained significant detection in all bins but the last.
We further used \textsc{ixpeobssim} \texttt{PHA1} algorithm to generate the weighted Stokes $I$, $Q$, and $U$ parameters with different binning in energy. For the analysis, we used 11 bins with the bin size 0.5\,keV in 2--7\,keV, and 1\,keV for the last 7--8\,keV bin, respectively, while for plotting, we used 5 bins to compare with the unweighted \texttt{pcube} results. We further generated $Q/I$ and $U/I$ spectra that were converted to \textsc{xspec}-employable FITS spectra with the \textsc{ftool} \texttt{flx2xsp} with unit response matrices. 
%The $I$, $Q$,  $U$, $Q/I$, and $U/I$ were binned to 11 bins with the bin size 0.5\,keV in 2--7\,keV, and 1\,keV for the last 7--8\,keV bin, respectively.
% Data binning - describe
%The data were binned to 5 bins to have significant detection in each bin. 

\subsection{NICER}

The Neutron-Star Interior Composition Explorer (NICER) is a soft X-ray timing mission deployed on the International Space Station (ISS).  NICER is composed of 52  silicon-drift detectors, sensitive from $\sim$0.2--12~keV, with $<100$~ns timing fidelity \citep{Gendreau2012}.  Each detector is paired with a single-bounce reflector optic, all mutually co-aligned on the sky.  NICER carried out two observations of \mbox{LMC~X-3} during the IXPE campaign, on 2023 July 8 and 17, corresponding to OBSIDs 6101010117 and 6101010118, respectively.  These observations were obtained after one of the detector thermal shields was damaged in May 2023, which resulted in a light leak during ISS daytime which results in optical loading of the detectors, producing an increase in noise and potential packet losses.\footnote{https://heasarc.gsfc.nasa.gov/docs/nicer/analysis\_threads/light-leak-overview/}  

\mbox{LMC~X-3} could only be observed during ISS daytime, and as a result, we found it necessary to use nonstandard filtering to recover usable data. For both observations, 42 of NICER's 52 detectors were turned on. We screened the active detectors for outlier behavior on the basis of rates of X-ray, overshoot, and undershoot events, flagging $>10$-(robust) $\sigma$ outliers from the detector ensemble.  This resulted in discarding all data from between 1--7 detectors per continuous GTI interval. We obtain a useful exposure time of $\approx$2.2~ks and $\approx$160~s, for the two observations, respectively, and were initially extracted separately per continuous GTI segment. After checking their mutual consistency, we merged them together for the spectral analysis.
In order to avoid contamination from low-energy noise events which are exacerbated by the light leak, we restrict our analysis to an energy range $>0.5$~keV. Response products were generated based on the number of active detectors, and rates were adjusted for $\gtrsim 1\%$ detector deadtime caused primarily by optical-loading events.

Because NICER is non-imaging, the spectral background is determined via various empirical models \citep{Remillard_2022}. We adopt the \texttt{SCORPEON} background model,\footnote{\url{https://heasarc.gsfc.nasa.gov/docs/nicer/analysis_threads/scorpeon-overview/}} and is normalized to the number of selected detectors.

\subsection{NuSTAR}

Three accompanying observations by the {\nustar} satellite \citep{Nustar_2013} were performed at the beginning, in the middle and at the end of the  {\ixpe} observation with the total net exposure time of $\approx$85\,ks, see more details in Table~\ref{table:list_observations}. 
%, on 2023 July 9 with the net exposure time of 27.6\,ks (ObsID: 30902041002), on July 14 with the net exposure time of 28.1\,ks (ObsID: 30902041004), and on July 19--20 with the net exposure time of 29.0\,ks (ObsID: 30902041006).
\nustar data were reduced using the standard Data Analysis Software (\texttt{NuSTARDAS}). The \nustar calibration files  available in the CALDB database were used to calibrate the cleaned event files, produced by the \texttt{nupipeline} task.
The source regions were selected as circles with a radius of 60\arcsec\ centered on the source image, and  background regions with radius 90\arcsec\ were selected from the corner of the same quadrant in the source-free region.
The source spectrum is soft and background dominates over source above 20\,keV. Therefore, we limit the \nustar data of \mbox{LMC X-3} at high energy to be below 20\,keV in all spectral analysis.

\subsection{Neil Gehrels \swift Observatory}

Simultaneous to {\nustar}, the Neil Gehrels \swift Observatory \citep{Gehrels2004} observed \mbox{LMC~X-3} with exposure times of 1.3 ks (ObsID: 00089714001, 2023 July 09), 1.8 ks (ObsID: 00089714002, 14 July 2023), and 1.9 ks (ObsID: 00089714003, 2023 July 20). Individual X-ray Telescope (XRT; \citet{Burrows2005}) spectra were extracted using the standard online tools provided by the UK Swift Science Data Centre \citep{Evans2009} using a source region of 60\arcsec\ radius centered at the source location. 

\subsection{Software tools for data analysis}
We used \textsc{heasoft}\footnote{https://heasarc.gsfc.nasa.gov/docs/software/lheasoft/} software package \citep{Heasoft1995}, version 6.31, for the data reduction and rebinning, and \textsc{xspec} software package \citep{Arnaud1996}, version 12.13, for the spectral analysis. We rebinned all the {\nicer}, {\nustar}, and {\swift} data with the tool \texttt{ftgrouppha} and applied optimal binning \citep{Kaastra2016} together with the condition for a minimum signal-to-noise ratio to be equal to 3. For the time-averaged spectral fits, we combined observations from multiple exposures using \textsc{addspec.py}.\footnote{https://github.com/JohannesBuchner/BXA} Across the entire paper, the errors are quoted as 90\% confidence levels if not stated otherwise. 
%The cosmological parameters used are H$_0$=70, $q_0=0$, $\lambda_0=0.73$.

%\begin{figure}
%    \centering
%    \includegraphics[width=0.49\textwidth]{ LightCurves.png}
%    \caption{X-ray light curves of LMC X-C. Black solid dots: IXPE light curve for the energy range 2–8 keV. Blue solid dots: NICER light curve for the energy range 1–12 keV. Green solid dots: NuSTAR light curve for the energy range 3–78 keV from the instrument A of NuSTAR.  Red solid dots: \swift light curve for the energy range 0.2–10 keV.}
%    \label{fig:lightcurves}
%\end{figure}

\begin{figure}
    \centering
    \includegraphics[width=0.49\textwidth]{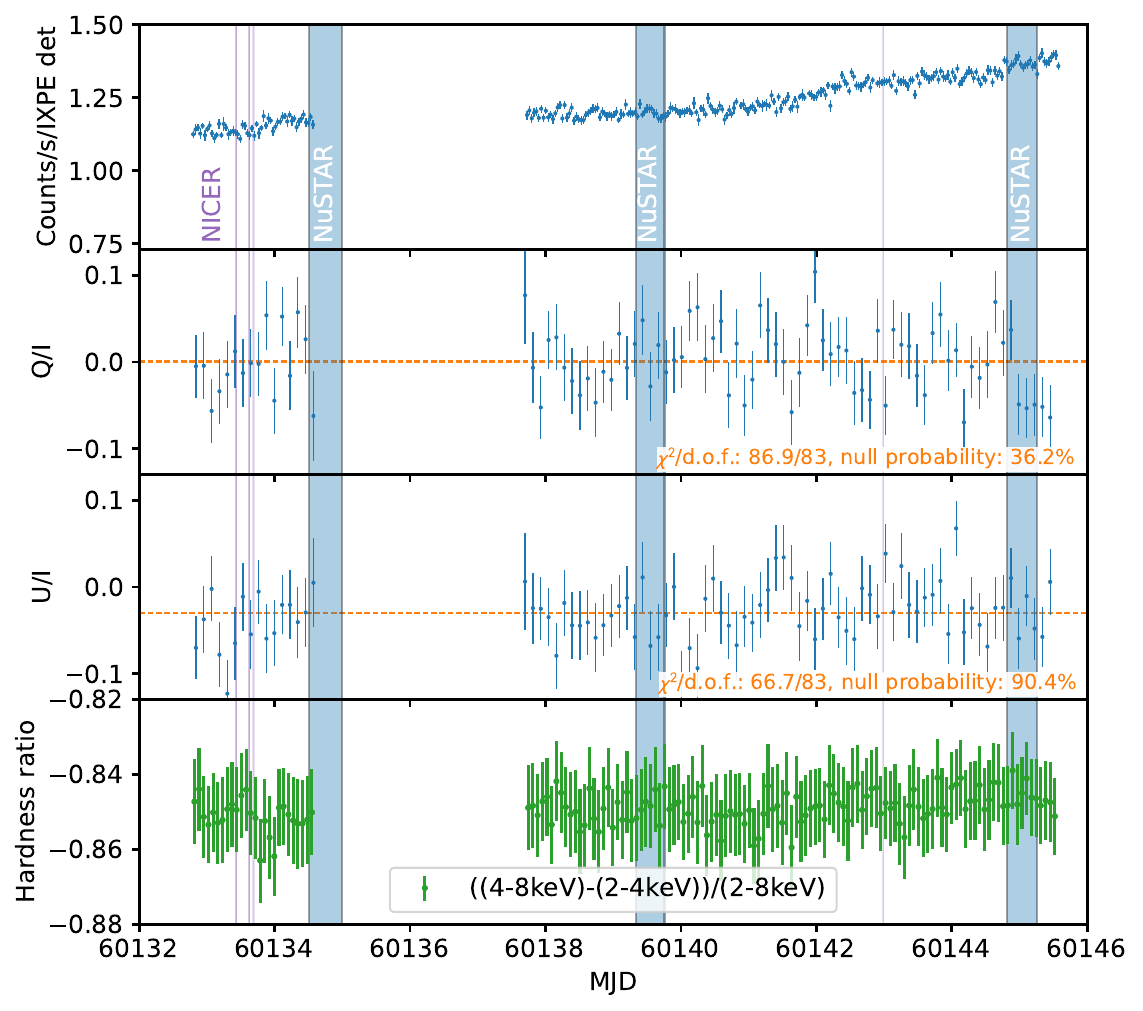}
    \caption{{\ixpe} light curve: variation of the counts ({\em top} panel), normalized Stokes parameters $Q/I$ ({\em second} panel) and $U/I$ ({\em third} panel), and spectral hardness ({\em bottom} panel), as a function of time (in Modified Julian Date). The dates of the accompanying observations by {\nicer} and {\nustar} are indicated as shaded regions.}
    \label{fig:stokes_orbit}
\end{figure}

\begin{figure*}
    \centering
    \includegraphics[width=0.49\textwidth]{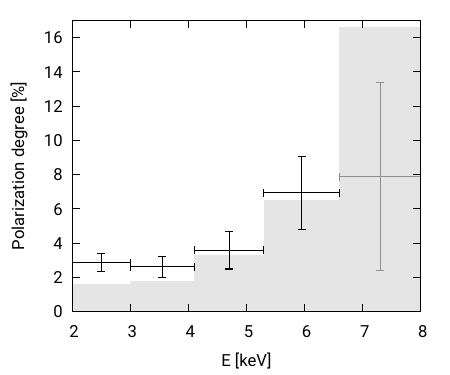}
    \includegraphics[width=0.49\textwidth]{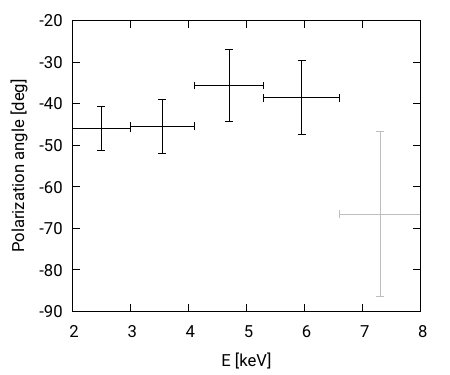}
    \caption{Measured  PD ({\it left}) and  PA ({\it right}) shown with 1$\sigma$ error bars. The shaded area in the PD-plot is an estimate of the MDP$_{99}$, showing the significant polarization measurements from 2\,keV up to $\approx 6.5$\,keV.}
    \label{fig:pdpa}
\end{figure*}

\section{Results} \label{Results}

\subsection{X-ray polarimetric measurements}
\label{polarization-pcube}
%\begin{itemize}
%    \item IXPE light curve - counts, hardness, Q, U
%    \item Q,U, polarization fraction and angle
%\end{itemize}

The IXPE light curve is shown in Figure~\ref{fig:stokes_orbit}. \mbox{LMC~X-3} shows a steady continuous increase of the flux during the exposure. The count rate (averaged over all 3 GPDs) increased from about 1.2 cts\,s$^{-1}$ at the beginning of the observation to 1.5 cts\,s$^{-1}$ at the end of the observation. The polarization is characterized by the Stokes $Q$ and $U$ parameters divided by the Stokes parameter $I$ (total number of counts). The $Q/I$ and $U/I$ light curves are shown in the middle panels of Figure~\ref{fig:stokes_orbit}.  The variations observed are consistent with statistical fluctuations only, with a successful joint fit achieving  $\chi^2$/dof=153.7/166. There is also no evidence for any significant changes in the spectral hardness, defined as the difference between counts in the hard (4--8\,keV) and soft (2--4\,keV) energy bands divided by the total number of counts in the 2--8\,keV band, as shown in the bottom panel of  Figure~\ref{fig:stokes_orbit}.

Using the time-integrated measurements of the Stokes parameters, we derived the PD and PA using the \texttt{xpbin} tool within the \texttt{pcube} algorithm. The average 2--8\,keV PD for all three detectors is PD $= 3.1\% \pm 0.4\%$, and the PA $= -45\degr \pm 4\degr$ with the 1$\sigma$ errors. The measurements are above the so-called minimum detectable polarization, MDP$_{99}$, which is the degree of polarization, for which the probability of the detection of the corresponding amplitude modulation only by chance is 1\% \citep{Weisskopf2010}.
In our observation MDP$_{99}$ = 1.23\% in 2--8\,keV.
The energy dependence of PD and PA is shown in Figure~\ref{fig:pdpa}, with the data binned in 5 energy bins. %Significant detection of the polarization is achieved in the entire band with the exception of the highest-energy bin (6.5--8 keV) where the MDP$_{99}$ is higher than the actual measurement. 
Measurement of the polarization above the MDP$_{99}$ is achieved in the entire band with the exception of the highest-energy bin (6.5--8 keV) where the MDP$_{99}$ is higher than the actual measurement. 
In the 2--5\,keV energy band, the PD is around 3\%. An increasing trend of polarization with energy is apparent from the plot, but the measurement uncertainty gets significantly larger (see more in Section~\ref{Discussion}). The polarization angle is consistent with being a constant with only possible small deviations at higher-energy bins.

\begin{figure}
    \centering
    \includegraphics[width=0.49\textwidth]{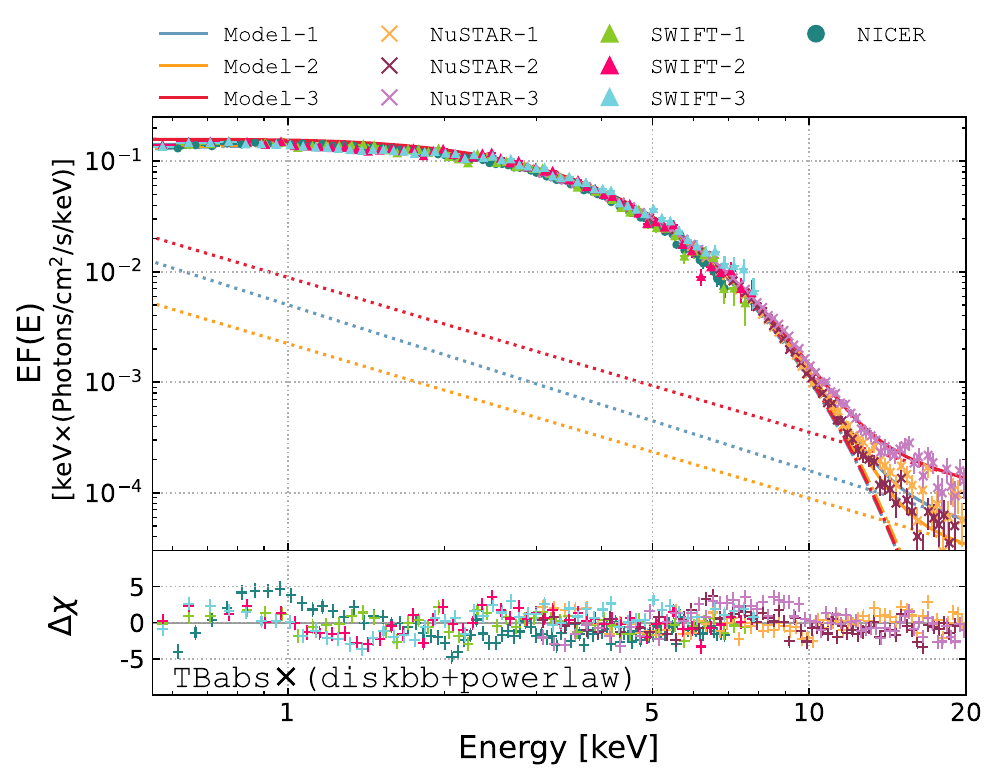}
    \caption{{\it Top:} Unfolded spectra ({\tt plot eufspec} in {\textsc{Xspec}}) with a joint spectral fit using the simple absorbed {\tt diskbb+powerlaw} model employing the {\nicer} and three {\swift} and {\nustar} exposures. {\it Bottom:} The residuals of the data from the model.}
    \label{fig:diskbbpo}
\end{figure}

%\begin{figure}
%    \centering
%    \includegraphics[width=0.47\textwidth]{3nuswini_diskbbpo_pllddelc.eps}
%    \caption{{\bf Top:} Joint spectral fit using the simple absorbed {\tt diskbb+powerlaw} model employing the {\nicer} and three {\swift} and {\nustar} exposures. {\bf Bottom:} The residuals of the data from the model.}
%    \label{fig:diskbbpo}
%    \label{fig:emo}
%\end{figure}

%\begin{figure}
%    \centering
%    \includegraphics[width=0.49\textwidth]{3nu_plemo.eps}
%    \caption{Model decomposition of the simple {\tt diskbb+powerlaw} fit of three {\nustar} exposures.}
%    \label{fig:emo}
%\end{figure}

%\begin{figure*}
%    \centering
%    \includegraphics[width=0.79\textwidth]{allspec_tbfeosimsli_edggab_pllddelc.eps}
%    \includegraphics[width=0.79\textwidth]{combined-compare-eeuf.png}
%     \caption{Alternative figure to Figure 4. {\bf Top:} {\nicer}, {\nustar}, {\swift} and {\ixpe} observations fitted by a best-fit spectral model using {\tt slimbh} and {\tt simpl} models. {\bf Middle:} The residuals of the data from the model using {\tt simpl*kerrbb}. {\bf Bottom:} The residuals of the data from the final best-fit model.}
%    \label{fig:bestfit}
%\end{figure*}

\subsection{X-ray spectral analysis}
%\begin{itemize}
%    \item simple analysis with diskbb+po.. - NICER combined + Swift + NuSTAR (all 3 exposures)
%    \item joined spectral analysis NICE+\swift+NuSTAR+IXPE 
%    \begin{itemize}
%        \item diskbb+po ?
%        \item kerrbb + nthcomp or simpl*kerrbb ?
%        \item simpl*slimbh
%    \end{itemize}
%\end{itemize}

To properly model and interpret the polarization measurements, a robust spectral fit is first needed. For this purpose, the long {\ixpe} observation was accompanied by several exposures with the sensitive instruments suitable for a broad-band spectroscopic analysis (see Table~\ref{table:list_observations}). 

\subsubsection{Time-resolved spectroscopy with a multi-color disk blackbody and power-law model}

We first analyzed the {\nicer}, {\nustar} and {\swift} spectra using a simple absorbed multi-colored disk blackbody emission \citep{Mitsuda1984} and power-law component for the Comptonization. We used {\tt tbabs} model to account for absorption in the line-of-sight in our Galaxy \citep{Wilms2000} and fixed the value of the hydrogen column density $N_{\rm H} = 4.5 \times 10^{20}$\,cm$^{-2}$ from a full sky HI survey \citep{HI4PI}. The model in \textsc{xspec} notation is {\tt tbabs*(diskbb + powerlaw)}. We further add a cross-normalization factor to account for changes between different instruments. The disk temperature as well as the power-law photon index values were linked between different instruments and also between different exposures. Only the normalization factors of both components (disk blackbody and power law) were allowed to vary to determine if there is any spectral variability and of which component. 

We found that the spectrum is dominated by the disk blackbody emission with the Comptonization component significant only for {\nustar} observations that have a coverage above 10\,keV, see Figure~\ref{fig:diskbbpo}. For the inner disk temperature, we obtained the value $kT \approx 1.1$\,keV. 
%$kT = 1.10 \pm 0.01$\,keV. 
We get the power-law photon index of 
%$\Gamma = 2.37 \pm 0.06$. 
$\Gamma \approx 2.4$. 
%From the 3 exposures of {\swift} and {\nustar} across the {\ixpe} observational campaign, we can study potential spectral variability of the source. 
From comparing the three {\swift} spectra, we see that there is no significant variability in the soft X-ray band, confirming the results suggested from the {\ixpe} hardness ratio shown in Figure~\ref{fig:stokes_orbit}. 
%The three {\swift} spectra are very similar, 
The only apparent but small difference is around 1.5\,keV between the first and the other  observations. Similar discrepancies below 2\,keV are likewise seen between {\swift} and {\nicer} measurements. These residuals lead to the fit that is not formally acceptable, with a chi-square value of $\chi^2 = 1255$ for 411 degrees of freedom ($\chi^2_{\rm red} \approx 3$).

The {\nustar} observations reveal a clear variability of the Comptonization component above $\approx 10$\,keV. The simple {\tt diskbb + powerlaw} model allows us to estimate the fraction that is the Comptonized emission, which is less than $1\%$ in 2--8\,keV. The model decomposed in the two components is plotted in Figure~\ref{fig:diskbbpo} for the three {\nustar} observations. The strongest Comptonization component is measured in the last observation. However, it is evident from the plot that the Comptonization contributes very little to the {\ixpe} 2--8\,keV energy band, and thus we can assume that the measured polarization is related to the main component, which is the thermal emission of the accretion disk. 

\begin{table*}[t]
\centering
\footnotesize
\caption{Spectral fit parameters with the final preferred spectral model.}
\begin{tabular}{ccccccc}
\hline
\hline
%\rule{0cm}{0.3cm}
Component & Parameter  & Description & \multicolumn{4}{c}{Value}\\
& (units) & &
{\nicer} & {\nustar} & {\swift} & {\ixpe} \\ 
%\hline
\hline
\rule{0cm}{0.3cm}
{\tt TBfeo} & $N_\textrm{H}$ ($10^{22}$\,cm$^{-2}$) & H column density &  \multicolumn{4}{c}{$0.046 \pm 0.003$} \\
\rule{0cm}{0.3cm}
& O & abundance &   \multicolumn{4}{c}{$0.3 \pm 0.2$} \\
& Fe & abundance &   \multicolumn{4}{c}{$0.7^{+0.4}_{-0.5}$} \\
\hline
\rule{0cm}{0.3cm}
{\sc{slimbh}} & $M_{\textrm{bh}}$ ($M_\odot$) & Black hole mass & \multicolumn{4}{c}{$6.98$ (frozen)}  \\ 
    & $a/M$ & Black hole spin & \multicolumn{4}{c}{$0.19 \pm 0.02$} \\
    &  $L_{\rm Edd}$ & Luminosity & $0.40 \pm 0.01$ & $0.43\pm 0.01$  & $0.41\pm 0.02$ & $0.45\pm 0.02$ \\
    & $i$ (deg) & Inclination & \multicolumn{4}{c}{$69.2$ (frozen)} \\
    & $\alpha$ & Viscosity & \multicolumn{4}{c}{$0.1$ (frozen)} \\
    & $D_\textrm{bh}$ (kpc) & Distance & \multicolumn{4}{c}{$49.59$ (frozen)} \\
    & hd & Color hardening  & \multicolumn{4}{c}{$-1$ (i.e. using TLUSTY)} \\
    & $l_\textrm{flag}$ & Limb-darkening & \multicolumn{4}{c}{$0$ (frozen)} \\
    & $v_\textrm{flag}$ & Self-irradiation & \multicolumn{4}{c}{$0$ (frozen)} \\
    & norm & normalization & \multicolumn{4}{c}{$1$ (frozen)}  \\
\hline
\rule{0cm}{0.3cm}
{\sc{simpl}} & $\Gamma$ & Photon index &  \multicolumn{4}{c}{$2.7 \pm 0.3$} \\
%\multicolumn{4}{c}{$2.5^{\rm pegged}_{-0.2}$} \\ 
& FracSctr & scattered fraction &  \multicolumn{4}{c}{$0.012^{+0.001}_{-0.002}$}\\
\hline
\rule{0cm}{0.3cm}
$\chi^2$ / dof & & & \multicolumn{4}{c}{$317/265 \approx 1.2$} \\
%\hline
\hline
\end{tabular}
\label{table:spectral_fit}
\begin{tablenotes}
\item {\bf{Note:}} The final model in the \textsc{xspec} notation is \texttt{constant*edge*gabs*tbfeo*simpl*slimbh}. Model parameters for cross-calibration and instrumental features are summarized in Table~\ref{table:crosscal}.
%The final model also included cross-calibration uncertainties between {\nicer}, {\nustar}, {\swift}, and {\ixpe} instruments, modeled by a constant factor in a range $0.9-1.1$ with $C_{\rm NICER} = 1$, $C_{\rm Nu-FPMA} = 1.18 \pm 0.01$, $C_{\rm Nu-FPMB} = 1.16 \pm 0.01$, $C_{\rm \swift} = 1.23 \pm ..$, $C_{\rm IXPE-GPD1} = 0.86\pm 0.02$, $C_{\rm IXPE-GPD2} = 0.85\pm 0.02$, $C_{\rm IXPE-GPD3} = 0.8$ (pegged), and the gain correction was applied to response files for the {\ixpe} GPD detectors. The obtained gain parameters were: the slope %$g_{\rm s} = 1.017, 1.019$, and $1.017$ and the offset $g_o = -0.095, -0.13$, and $-0.11$\,keV for GPD 1, 2, and 3, respectively.
%$g_{\rm s} = 0.989 \pm 0.005, 0.990 \pm 0.006$, and $0.988$ and the offset $g_o = -0.04 \pm 0.02, -0.06 \pm 0.02$, and $-0.05 \pm 0.02$\,keV for GPD 1, 2, and 3, respectively.
%We further included a narrow Gaussian absorption model with $E_{\rm g}=2.06 \pm 0.04$\,keV, $\sigma_{\rm g}=0.10^{\rm pegged}_{-0.06}$, and strength $N_{\rm g}=0.13 \pm 0.05$ for {\nicer}, and $E_{\rm g}= 9.9 \pm 0.1 $\,keV and $N_{\rm g}=0.06 \pm 0.03$ for {\nustar}, and an instrumental edge $E_{\rm edge}=2.34 \pm 0.11$\,keV, $\tau_{\rm edge} = 0.007 \pm 0.004$ for {\nicer}.  \\
\end{tablenotes}
\end{table*}

\begin{table*}[t]
\centering
\footnotesize
\caption{Modeling cross-calibration and instrumental features in the final spectral fit.}
\begin{tabular}{ccccccccc}
\hline
\hline
%\rule{0cm}{0.3cm}
Component & Parameter & {\nicer}  & \multicolumn{2}{c}{\nustar}  & {\swift} & \multicolumn{3}{c}{\ixpe}  \\ 
\cline{4-5} \cline{7-9}
  &   &   & FPMA & FPMB &   & GPD~1 & GPD~2 & GPD~3 \\
%\hline
\hline
\rule{0cm}{0.3cm}
{\sc{constant}} & & $1$ (frozen) & $1.18 \pm 0.03$ & $1.16 \pm 0.03$  & $1.23 \pm 0.07$ & $0.86 \pm 0.02$ & $0.85 \pm 0.02$ & $0.80 \pm 0.02$\\
\rule{0cm}{0.3cm}
%\hline
%\rule{0cm}{0.3cm}
{\sc{gabs}} & $E$~(keV) & $2.07 \pm 0.04$ & \multicolumn{2}{c}{$9.9 \pm 0.1$} &  $2.14 \pm 0.04$ & \multicolumn{3}{c}{-}  \\ 
 & $\sigma$~(keV) & $0.08^{+0.02,\rm pegged}_{-0.06} $ & \multicolumn{2}{c}{$0.1$ (pegged)} &  $0.03 \pm 0.03$ & \multicolumn{3}{c}{-}  \\
{\sc{edge}} & $E$~(keV) & $2.35 \pm 0.05$ & \multicolumn{2}{c}{-} &  - & \multicolumn{3}{c}{-}  \\ 
 & $\tau$~(keV) & $0.02 \pm 0.02$ & \multicolumn{2}{c}{-} &  - & \multicolumn{3}{c}{-}  \\  
{\sc{gain}} & slope & - & \multicolumn{2}{c}{-} & - & $0.989 \pm 0.005$ & $0.990 \pm 0.006$ & $0.988 \pm 0.005$ \\ 
 & offset~(keV) & - & \multicolumn{2}{c}{-} & - & $-0.04 \pm 0.02$ & $-0.06 \pm 0.03$ & $-0.05 \pm 0.02$ \\ 
 %\\ 
\hline
\end{tabular}
\label{table:crosscal}
\end{table*}

\begin{figure}
    \centering
    \includegraphics[width=0.49\textwidth]{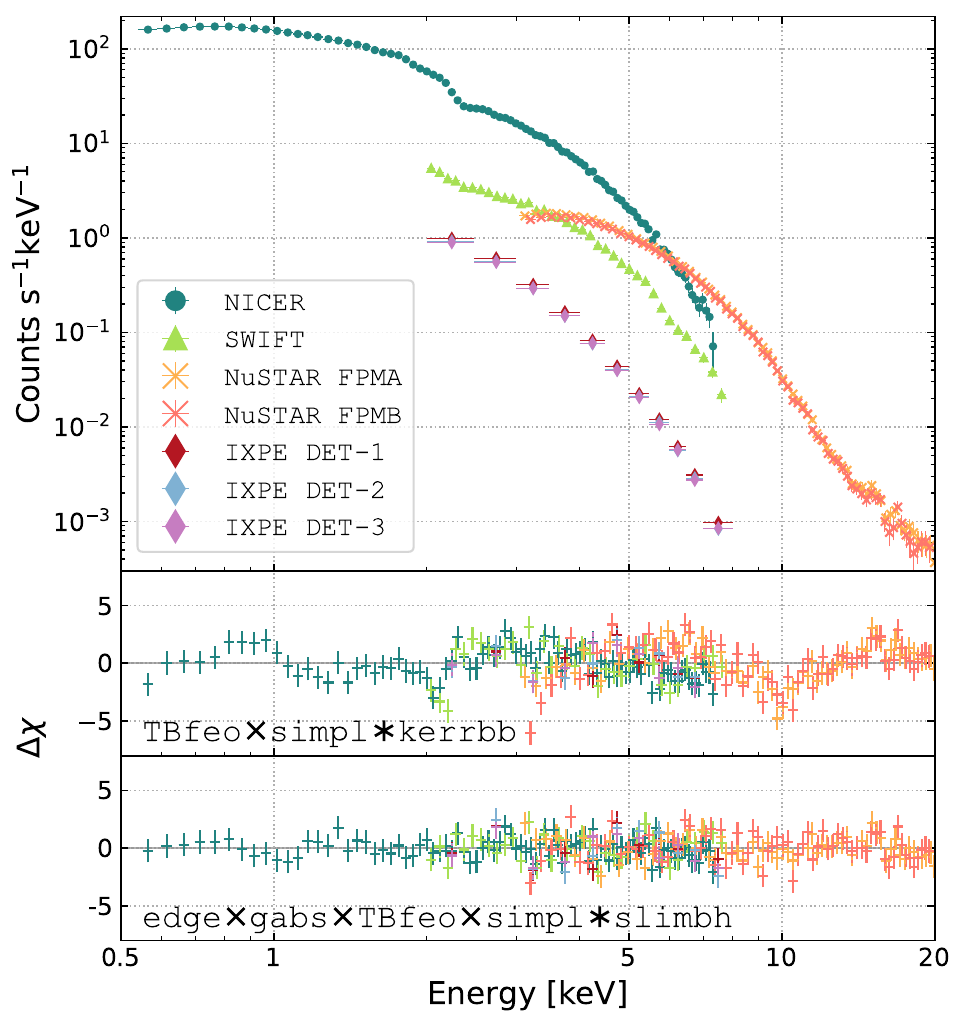}
     \caption{{\it Top:} Time-averaged {\nicer}, {\nustar}, {\swift}, and {\ixpe} data. % with the best-fit spectral model. %{\tt simpl * slimbh} 
     {\it Middle:} The residuals of the data from the model using {\tt tbfeo $\times$ simpl * kerrbb}. {\it Bottom:} The residuals of the data from the final best-fit model.}
    \label{fig:bestfit}
\end{figure}

%\subsubsection{Time-averaged spectral analysis}

\subsubsection{Spectral analysis with relativistic models for the accretion disk emission}

Owing to the detection of little spectral variability, we further use merged {\swift} and {\nustar} spectra. %over the three observations.
There are known calibration uncertainties with the {\ixpe} spectra, and we therefore applied the {\tt gain} model to fit the offset slope and intercept for {\ixpe}. %detector response matrices. 
We further performed a spectral analysis of 7 data sets (1 {\nicer}, 2 {\nustar} detectors FPMA and FPMB, 1 {\swift} and 3  detectors of {\ixpe}) with a cross-calibration constant fixed to 1 for {\nicer} and allowed to vary between 0.8 and 1.2 for the other detectors. Because of the discrepancies between \swift spectra below 2\,keV, we limit the merged {\swift} data to the 2--8\,keV energy band, and use only {\nicer} data below 2\,keV. Figure~\ref{fig:bestfit} shows the time-averaged spectra of different detectors in the first (top) panel.

%\subsubsection{Spectral analysis with relativistic models for the accretion disk emission}

We first employed a relativistic accretion disk model {\tt kerrbb} \citep{Li2005} convolved with a non-relativistic Comptonization model {\tt simpl} \citep{Steiner2009,Sunyaev1980}, allowing for both up- and down-scattering. %Based on previous constraints with the {\tt powerlaw} model, we set the photon index to be $\Gamma > 2$. 
For absorption, we employed the {\tt tbfeo} model allowing for different oxygen and iron abundances. Because the LMC is a low-metallicity environment, we let the abundances to be in the interval 0.25--1. The fit converged to $N_\textrm{H} \approx 0.03 \times 10^{22}$\,cm$^{-2}$ with the oxygen and iron abundances being pegged at their low-value limits at $0.25$.
%$0.25^{+0.05}_{\rm pegged}$ and $0.25^{+0.15}_{\rm pegged}$ of solar, respectively. 
We note that since the column density is lower than the column density expected in the line-of-sight in our Galaxy, the low oxygen and iron abundances may be an artifact of calibration uncertainties in the 0.5--1\,keV band for {\nicer} and/or due to variations of the absorption column within our Galaxy. No evidence for local absorption is consistent with \mbox{LMC~X-3}'s location at a large distance from the center of the LMC, away from any gaseous nebulae.

With the \texttt{tbfeo*kerrbb*simpl} model, we obtained the dimensionless black hole spin $a=0.20 \pm 0.02$, accretion rate $\dot{M} \approx (4.5-5.6) \times 10^{18}$\,g~s$^{-1}$, and the photon index pegged at $\Gamma = 2.0$, which was the lowest allowed value. We allowed the hardening factor of the \texttt{kerrbb} model to vary and we obtained $h_d \approx 1.7$ for {\nustar} 
%$h_d = 1.70^{+0.03}_{-0.10}$, $h_d = 1.87 \pm 0.01$
and $h_d \approx 1.9$ for {\nicer}. The fit was formally not acceptable with $\chi^2/{\nu} = 687/274 \approx 2.5$, mainly due to discrepancies between {\nicer} and {\nustar} data, whose residuals had opposite slopes in the overlapping energy band (see the second panel of Figure~\ref{fig:bestfit} at 3--8\,keV energy band). 

The lowest measured accretion rate $4.5 \times 10^{18}$\,g\,s$^{-1}$ corresponds to the luminosity $L = \eta \dot{M} c^2 \approx 0.3\,L_{\rm Edd}$, where $\eta$ is the accretion efficiency $\eta \approx 0.065$ (for $a=0.2$).
At such a luminosity, \mbox{LMC~X-3} might deviate from the standard thin disk model, and a slim disk scenario was proposed to take place at the high-luminosity regime \citep{Straub2011}. 

Therefore, we replaced the {\tt kerrbb} model with the {\tt slimbh} model \citep{Sadowski2011, Straub2011}, and we obtained a significantly better fit with $\chi^2/{\nu} = 414/277 \approx 1.5$. The {\tt slimbh} model improved the consistency of the data residuals between {\nicer} and {\nustar}. The only residuals were now narrow features around 2\,keV and 2.4\,keV for {\nicer} and 10\,keV for {\nustar}, which are visible in the second panel of Figure~\ref{fig:bestfit}.
Similar residuals were reported in some previous analyses and attributed to calibration uncertainties, see, e.g., \citet{Wang2021} for \nicer and \citet{Podgorny2023} for \nustar.
%and the residuals in the {\ixpe} spectra probably due to still imperfect calibration. 
For our final model, we account for the 2 and 10 keV features with narrow Gaussian absorption lines and the 2.4 keV feature with a `smeared-edge' component. The goodness of the final fit is $\chi^2/{\nu} = 319/266 \approx 1.2$, which without {\ixpe} data reduces to  $\chi^2/{\nu} = 259/243 \approx 1.06$, indicating that the remaining residuals might be due to cross-calibration discrepancies. The residuals from the best-fit model are shown in the third panel of Figure~\ref{fig:bestfit}.

The values of the best-fit model are summarized in Table~\ref{table:spectral_fit}. 
The spin value is consistent with the measurements using the {\tt kerrbb} model, $a \approx 0.20 \pm 0.02$. The spectral hardening in the {\tt slimbh} model is not a free parameter, but is instead calculated using the vertical structure computed using the TLUSTY code \citep{Hubeny1995}. The estimated luminosity is in the range $L = 0.40-0.45\,L_{\rm Edd}$ depending on which detector is considered (the slightly higher value for \ixpe can be, however, affected by the fitted cross-calibration constants lower than 1). The parameters of the Comptonization model {\tt simpl} were constrained well only from the {\nustar} spectra and therefore, we linked the values between the different detectors. The photon index is %$\Gamma = 2.5^{\rm pegged}_{-0.2}$
$\Gamma = 2.7 \pm 0.3$
and scattering fraction is $0.012^{+0.001}_{-0.002}$. Similarly, absorption was best constrained from the {\nicer} data and we linked the absorption parameters for the different detectors to it.

In our preferred model, the black hole mass and inclination are initially fixed to the values from the dynamical measurements \citep{Orosz2014}. Because in the IXPE observation of \mbox{Cyg X-1} the inclination of the innermost accretion disk from the X-ray spectroscopy and polarimetry was found to be different from the value for the orbital inclination \citep{Krawczynski2022}, we also performed an alternative spectral fit with free inclination. The best-fit value of the inclination changed slightly to $i = 71.8^{+2.0}_{-1.2}$\,deg, and corresponding inferred luminosity increased from $L = 0.43\,L_{\rm Edd}$ to $L =0.48\,L_{\rm Edd}$. The goodness of the fit improved by $\Delta \chi^2 = 310-317 = -7$ compared to the initial fit. This improvement was only marginal and we conclude that the inclination of the accretion disk constrained from the X-ray spectra is consistent with the inclination of the binary system constrained from the optical measurements.

\begin{figure}
    \centering
\includegraphics[width=0.49\textwidth]{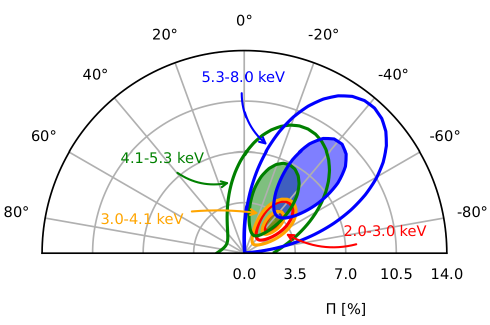}
    \caption{Polar plot of the polarization measured in different energy ranges with the spectral best-fit model. The filled contours correspond to the $68\%$ (1$\sigma$) and the outer contours to the $99.9\%$ confidence levels, respectively.}
    \label{fig:polar_plot_slimbh}
\end{figure}

\subsection{Spectro-polarimetric analysis}
\label{sect:spectropolarimetric}
%.. spectro-polarimetric fit
%\begin{itemize}
%    \item with final slimb spectral model
%    \item  with KYNBBRR
%\end{itemize}

We first included the {\ixpe} $Q$ and $U$ spectra into our analysis by taking our best-fitting spectral model from Table~\ref{table:spectral_fit} and assigning a constant PD and PA to it using the {\tt polconst} model. 
%We fixed the spectral parameters to the values obtained from the multi-observatory spectral fit presented in Table~\ref{table:spectral_fit}.
For $Q$ and $U$ spectra, we applied the same gain as for $I$ spectra, and we also kept the cross-normalization factors. The only free parameters were the PD and PA, noted as $A$ and $\psi$ in the {\tt polconst} model. Considering the full {\ixpe} bandpass yields a PD of $A = 3.2\% \pm 0.6 \%$ and a PA  $\psi = -42 \degr\pm 6\degr$. 

To investigate any energy dependence, we performed the fit in four different energy bands spanning 2--8 keV and calculated contours using 50 steps in each parameter. 
%Compared to Figure~\ref{fig:pdpa}, we merged the two highest energy bins in this analysis to have significant measurements in every bin.
We defined the energy bands as: 2--3, 3--4.1, 4.1--5.3, and 5.3--8\,keV.
Figure~\ref{fig:polar_plot_slimbh} shows the resulting contours in the polar plot of PD and PA. While there is an apparent trend of increasing PD with the energy as in Figure~\ref{fig:pdpa}, the significance of the change is not high. The PD is consistent with being a constant below 5\,keV, and increases with energy above 5\,keV only at $1\sigma$ confidence level.

We then proceeded to fit physical models that self-consistently predict energy-dependent polarization properties. Since no polarized slim-disk model currently exists, we were limited to fitting standard thin-disk models. We employed models from the relativistic package \textsc{ky} \citep{Dovciak2004}, which has been developed for spectral, timing, and polarimetry analysis. We employed the relativistic thin Novikov-Thorne disk model \kynbbrr \citep{Taverna2020}. This includes the same parameters as {\tt kerrbb}, with some extra parameters required to specify the polarization properties such as the position angle of the disk (and black hole) rotation axis on the plane of the sky $\chi_0$, the optical depth of electron scattering in the disk atmosphere $\tau$, and the albedo that defines the reflectivity of the disk surface for returning radiation (radiation that is lensed by a black hole such that it returns and reflects on the other part of the disk before reaching an observer). The {\tt Stokes} parameter in the model allows us to define how the polarization is calculated. We used {\tt Stokes} = 1 that reads the polarization from loaded $I$, $Q$, and $U$ spectra. The polarization characteristics of this model for various sets of accretion disk parameters and a more-detailed model description are presented in \citet{Mikusincova2023}.

We also employed the absorbed \kynbbrr model using the spectral parameters from the fit with the {\tt tbfeo$\times$ simpl*kerrbb} model. 
To avoid the fit being dominated by the total count spectrum, we fitted only $Q$ and $U$ spectra (i.e., without $I$) with free $\chi_0$  and additionally a free normalization of the model to account for differences between \kynbbrr and {\tt kerrbb}.
We obtained an acceptable fit with $\chi^2/{\nu} = 28.4/28 \approx 1.0$ with $\chi_0 = 47\degr \pm 6\degr$ and normalization parameter $N_{\rm K} = 0.066 \pm 0.013$.\footnote{\kynbbrr does not have distance as a model parameter. The normalization would be around 1 for the source at the distance of 10\,kpc. The distance of \mbox{LMC~X-3} is $\approx 50$\,kpc and thus, the normalization value is expected to be around $1/5^{2} \approx 0.04$, roughly consistent with the measurements.}

\begin{figure}
    \centering
    \includegraphics[width=0.49\textwidth]{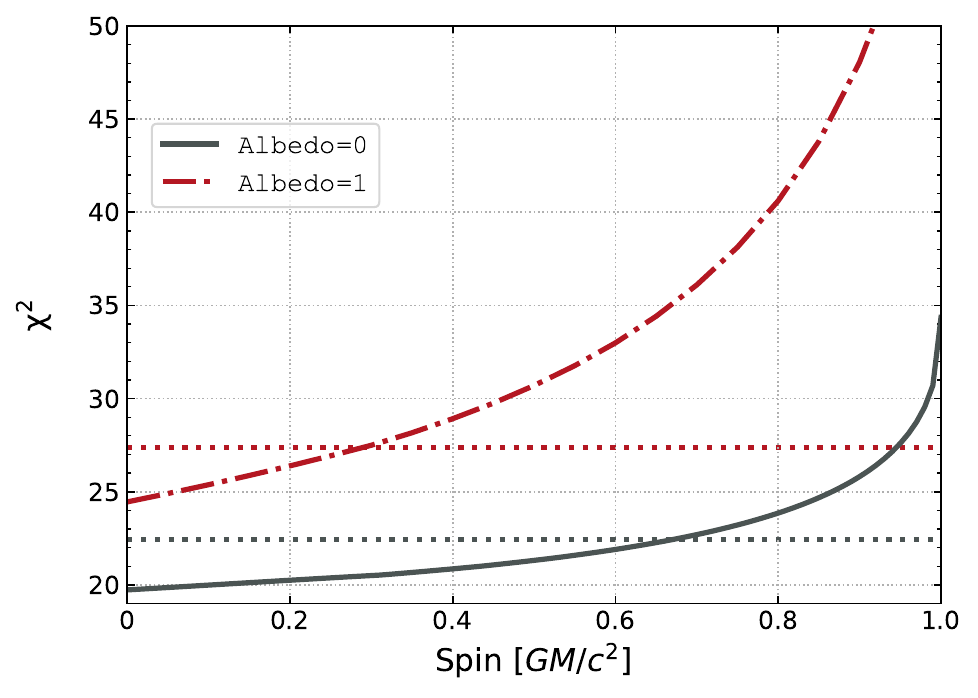}
    \caption{Polarimetry constraints of the black hole spin, expressed as goodness of the fit (as the $\chi^2$ values) versus black-hole spin from $Q/I$ and $U/I$ fitting with the {\kynbbrr} model for two extreme values of albedo: 0 (black, solid line) and 1 (red, semi-dashed line). The dotted horizontal lines represent 90\% confidence levels.}
    \label{fig:qnun_spin_fit}
\end{figure}

\begin{figure*}
    \centering
    \includegraphics[width=0.49\textwidth]{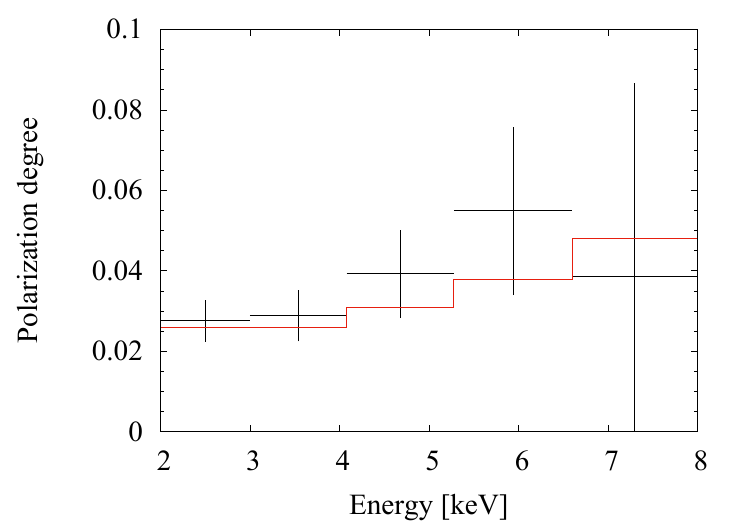}
    \includegraphics[width=0.49\textwidth]{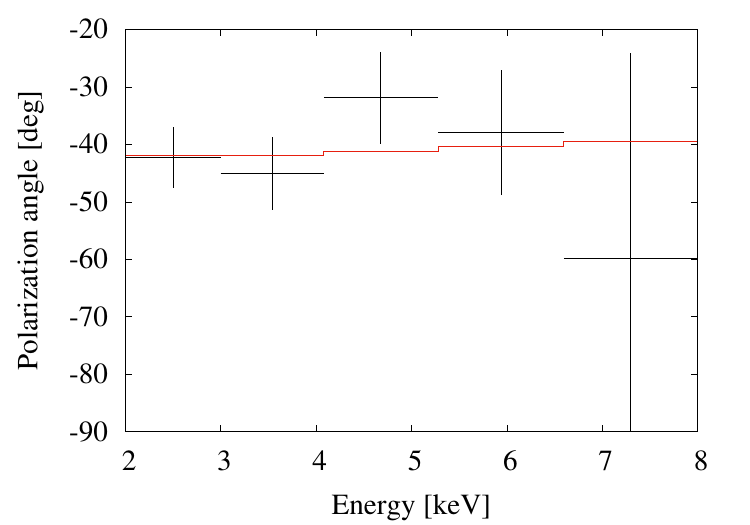}
    \caption{PD ({\it left}) and PA ({\it right}) fitted by the {\kynebbrr} model with the optical depth $\tau = 4.5$.}
    \label{fig:ixpe_kynebbrr}
\end{figure*}

\subsection{Black hole spin measurements from the polarimetry}

For fitting the black-hole spin from the polarimetric measurements only, independently of the total spectrum, we employed the normalized $Q/I$ and $U/I$ spectra, to which we applied the \kynbbrr model with the normalization fixed to 1. The \texttt{Stokes} parameter in the \kynbbrr model needs to be set to 8 for $Q/I$ and 9 for $U/I$, respectively. We tested two cases of albedo, 0 and 1. While the albedo = 0 means that no returning radiation is taken into account, albedo = 1 corresponds to the 100\% reflectivity of the gravitationally light-bended returning radiation. The albedo is important mainly for a highly spinning black-hole when the ISCO extends closer to the black hole and more returning radiation is expected \citep{Cunningham1976}.

For the albedo equal to 0, we fitted the $Q/I$ and $U/I$ spectra with the black-hole spin $a$ and $\chi_0$ as free parameters. We obtained a perfectly acceptable fit with $\chi^2/{\nu} = 19.9/20 \approx 1.0$, $a < 0.66$ and $\chi_0 = 44\fdg6 \pm 6\fdg4$. For a non-zero value of the albedo, the model is currently calculated for 20 values of the spin and does not allow for a direct fitting of this parameter. We applied the \texttt{steppar} command in \textsc{xspec} to calculate the $\chi^2$ values for the different values of the black-hole spin, and for the comparison, we performed the same procedure for albedo equal to 0.

Figure~\ref{fig:qnun_spin_fit} shows the dependence of the fit goodness against the spin value for the two cases with albedo equal to 0 and 1, respectively. 
With the 90\% confidence, the black-hole spin is required to be lower than $0.66$ % a<0.66
for albedo = 0 and lower than 0.3 for albedo = 1. The case of albedo = 0 is preferred by the fit with lower $\chi^2$ values. The results indicate that the sole polarimetry measurements are consistent with the low black-hole spin in \mbox{LMC~X-3}, independently of the spectral fitting.

\section{Discussion and conclusions} \label{Discussion}

With the best-fit spectral model, we obtained for the polarimetry: PD $= 3.2\% \pm 0.6 \%$ and PA$= -42\degr \pm 6\degr$ in the 2--8\,keV energy band, assuming a constant polarization over energy. These measurements are in perfect agreement with the values (PD $= 3.1\% \pm 0.4 \%$ and PA$= -45\degr \pm 4\degr$) obtained from an alternative analysis using the \texttt{xpbin} tool within the \texttt{pcube} algorithm (see Section~\ref{polarization-pcube}).

%some discussion on PA - no radio and optical constraints come here..
Regarding the PA measurements, there are no known large-scale physical structures associtated with \mbox{LMC~X-3} with which to compare the PA. \mbox{LMC~X-3} is persistently in the high/soft state and no jet has been detected in the radio despite several efforts \citep{fender_radio_1998, gallo_universal_2003,Lang_2007}. 
%Therefore, in order to put constraint on the PA orientation, we looked for an ionization cone along the rotational axis of the black hole. 
There is also no evidence for ionization cones in the far-UV \citep{2003AJ....126.2368H}, %, with the Far UV Spectroscopic Explorer (FUSE), and 
or presence of any significant emission or absorption lines in soft X-rays \citep{Page_2003}.
% with XMM-Newton, the continuum dominates the spectra without. 
%Simulations of \mbox{LMC~X-3} from \cite{nazarenko_formation_2008} mention a low-density tunnel along the rotational axis of the black hole, with an estimated opening angle between 3 to 5 degrees. But no observations have yet confirmed these results.

The level of the PD is consistent with expectations for the thermal disk emission around a black hole with a low spin and high inclination (see the case of $a=0$ and $i=70\degr$ in Figure~4 in \citet{Mikusincova2023} with a constant PD just slightly below 3\%). The Novikov-Thorne model, assumed in the \kynbbrr spectral model, provides a perfect fit to the X-ray polarimetry despite the best-fit spectral model employing a slim disk. This is most likely explained by the limited statistics of the X-ray polarimetry fit, while the spectral fit is sensitive to small differences between the \kerrbb and \texttt{slimbh} models, which can possibly be attributed to the treatment of the spectral hardening in these models. 

The black-hole spin, solely constrained from the X-ray polarimetry using normalized Stokes parameters $Q/I$ and $U/I$, is consistent with the results of determining the spin from spectral fitting. The constraints on the spin are tighter if the reflection of the returning radiation is taken into account, but even for a model with no reflected returning radiation (albedo=0), the spin is constrained to be less than 0.7 (see Figure~\ref{fig:qnun_spin_fit}).

While the X-ray polarization is well described by the model with constant PD with energy, the data tentatively indicate an increase of the PD with energy, which is most prominently visible in the plot showing the results from the {\tt pcube} analysis (see Figure~\ref{fig:pdpa}, especially above 5\,keV). At higher energies, however, the uncertainty of the polarization measurements increases due to a lower number of counts and thus in our case, the PD increase with energy is suggested with only marginal statistical significance (see Figure~\ref{fig:polar_plot_slimbh}).

%The energy-averaged measurement is dominated by the higher-significance detection with a higher count rate at low-energy bins. However, the data tentatively indicate an increase of polarization degree with energy, which is most prominently visible in the plot of the polarization degree from the {\tt pcube} analysis, see Figure~\ref{fig:pdpa}. However, a more detailed analysis using the best spectral fit shows the polarization fraction to be consistent with being constant in the 2--5\,keV energy band. At higher energies, the uncertainty of the polarization measurements increases due to a lower number of counts and thus in our case, the PD increase with energy is suggested but with marginal statistical significance (see Figure~\ref{fig:polar_plot_slimbh}). 

The increasing energy dependence is included in the model variant {\kynebbrr} of the KY package. We employed this model using the same spectral parameters from the fit of the {\tt simpl*kerrbb} model, and we fitted only $Q$ and $U$ spectra. The parameter $\tau$ was a free parameter together with $\chi_0$ and the normalization of the model. We fixed the albedo to 0, and we obtained a very good fit with $\chi^2/{\nu} = 26.5/27 \approx 1.0$. The best-fit parameters are $\chi_0 = 47\degr \pm 6\degr$, $\tau = 4.5^{+4.5}_{-2.6}$, and $N_{\rm K} = 0.04 \pm 0.02$. The PD and PA using this model are shown in Figure~\ref{fig:ixpe_kynebbrr}. However, compared to results obtained with {\kynbbrr} (see Section~\ref{sect:spectropolarimetric}, with $\chi^2/{\nu} = 28.4/28$), the improvement is not statistically significant.

\begin{figure}
    \centering
    \includegraphics[width=0.49\textwidth]{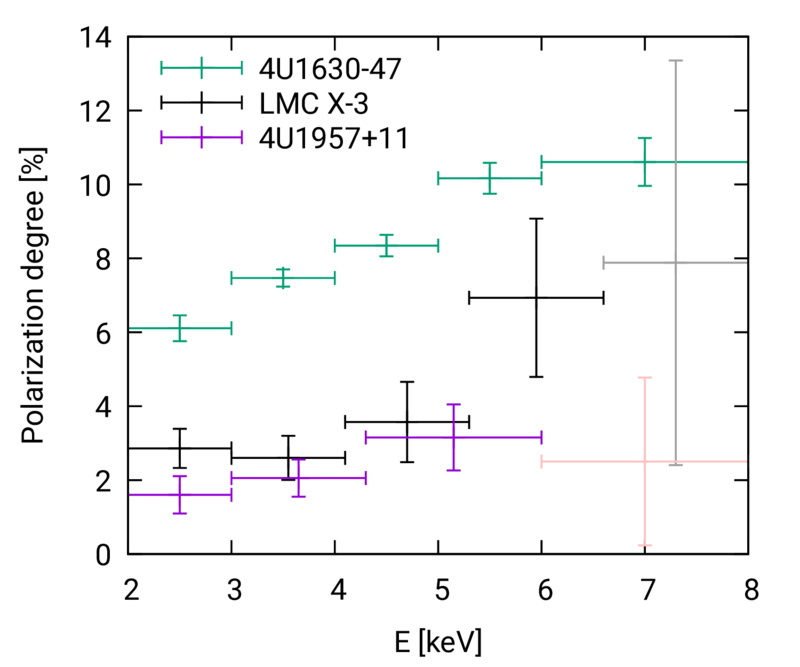}
    \caption{Comparison of the energy dependence of the PD in X-ray binaries in the high/soft state dominated by the thermal emission of the accretion disk.}
    \label{fig:pd_comparison}
\end{figure}

The trend of increasing PD with energy and roughly constant PA seem, nevertheless, to be  typical for observations of BHXRBs in the high/soft state. A comparison of the behavior of the PD with energy for three BHXRBs in the high/soft state is shown in Figure~\ref{fig:pd_comparison}, where a similar trend is apparent for all sources. Various explanations have been proposed, and different scenarios might be responsible for the observed trend in  different sources. In the most prominent case of 4U 1630$-$47, the PD increase with energy is statistically very significant. A possible explanation attributes this to absorption in the upper layer of the accretion disk in combination with a relativistic bulk motion, which could also explain a higher PD than expected from an accretion disk with the electron scattering dominated atmosphere \citep{Ratheesh2023}. For 4U 1957+11, the increase can be explained by a combination of a high spin value and high albedo (Marra et al., in prep.). High albedo is, however, unlikely in the case of \mbox{LMC~X-3} given the low value of the black-hole spin.
To perform a robust statistical test of the significance of the PD's increase with energy, a longer observation would be needed to obtain a significant measurement up to 8\,keV.

%While the first case with no returning radiation is more appropriate for very low spin values, the albedo equal to 1, i.e. assuming 100\% reflectivity of the returning radiation, would be more realistic for higher spin values. 

\section{Summary} \label{Summary}

We report on the first X-ray polarimetric observation of the accreting stellar-mass black hole \mbox{LMC~X-3} with the IXPE. 
The polarization is significantly detected with the PD being $3.2\% \pm 0.6 \%$ and PA measured as $-42\degr \pm 6\degr$ in the 2--8\,keV energy band.
We performed a spectro-polarimetric fit including the accompanying observations by {\nicer}, {\nustar}, and {\swift} satellites, showing that the X-ray spectrum is best modeled by a slim accretion disk with the intrinsic luminosity $L \approx 0.4\,L_{\rm Edd}$. We used the spectral data to measure the black hole spin $a \approx 0.2$. 
%consistent with the previous spectroscopic measurements. 
%, which is consistent with the thermal emission from a highly inclined ($\approx 70\degr$) accretion disk around a low-spin black hole. 
Using solely the polarimetric normalized Stokes parameters $Q/I$ and $U/I$, we obtained for the black-hole spin: $a < 0.7$ at 90\% confidence level, in agreement with the spectroscopic measurements.
%from the north direction. 

%% Also note that the acknowledgement environment does not support long amounts of text. If you have a lot of people and institutions to acknowledge, do not use this command. Instead, create a new 

\section*{Acknowledgments}

%\begin{acknowledgments} 

The \textit{Imaging X ray Polarimetry Explorer} ({\em IXPE}) is a joint US and Italian mission.  The US contribution is supported by the National Aeronautics and Space Administration (NASA) and led and managed by its Marshall Space Flight Center (MSFC), with industry partner Ball Aerospace (contract NNM15AA18C).  The Italian contribution is supported by the Italian Space Agency (Agenzia Spaziale Italiana, ASI) through contract ASI-OHBI-2017-12-I.0, agreements ASI-INAF-2017-12-H0 and ASI-INFN-2017.13-H0, and its Space Science Data Center (SSDC) with agreements ASI-INAF-2022-14-HH.0 and ASI-INFN 2021-43-HH.0, and by the Istituto Nazionale di Astrofisica (INAF) and the Istituto Nazionale di Fisica Nucleare (INFN) in Italy.  This research used data products provided by the {\em IXPE} Team (MSFC, SSDC, INAF, and INFN) and distributed with additional software tools by the High-Energy Astrophysics Science Archive Research Center (HEASARC), at NASA Goddard Space Flight Center (GSFC). This work made use of data supplied by the UK Swift Science Data Centre at the University of Leicester.

Ji\v{r}\'{i} S., M.D., Jak.Pod., Mic.Bur. and V.K. thank GACR project 21-06825X for the support and institutional support from RVO:67985815. 
A.Y. and Mai.Bri. acknowledge the support from GAUK project No. 102323. 
A.I. acknowledges support from the Royal Society.
A.V. thanks the Academy of Finland grant 355672 for support.
P.-O.P. acknowledges financial support from the French Space Agency (CNES) and the French High Energy National Programme (PNHE) of CNRS.
NRC, MRM, HK, KH, and SC acknowledge support by NASA grants 80NSSC22K1291, 80NSSC23K1041, and 80NSSC20K0329.

%\end{acknowledgments}


\vspace{5mm}
\facilities{{\ixpe}, {\nicer}, {\nustar}, {\swift}}


%\software{astropy \citep{2013A&A...558A..33A,2018AJ....156..123A},  
%          Cloudy \citep{2013RMxAA..49..137F}, 
%          Source Extractor \citep{1996A&AS..117..393B}
%          }



%\appendix


\bibliography{references_ixpeLMCx3}{}
\bibliographystyle{aasjournal}



\end{document}